\documentclass[prd,twocolumn,preprintnumbers,superscriptaddress,nofootinbib] {revtex4}
\pdfoutput=1
\usepackage[english]{babel}
\usepackage{amsmath,amssymb,amsfonts, bm,bbm,slashed, subdepth}
\usepackage{graphicx}
\usepackage[sort&compress]{natbib}
\usepackage{xcolor}
\usepackage{hyperref}
\usepackage{cleveref}
\usepackage{enumerate}
\usepackage{epsfig, subfigure}
\usepackage{setspace}
\usepackage{booktabs, tabularx}
\usepackage{units}

\newcommand{\be}{\begin{equation}}
\newcommand{\ee}{\end{equation}}
\newcommand{\ba}{\begin{array}}
\newcommand{\ea}{\end{array}}
\newcommand{\bea}{\begin{eqnarray}}
\newcommand{\eea}{\end{eqnarray}}
\newcommand{\balg}{\begin{align}}
\newcommand{\ealg}{\end{align}}
\newcommand{\bit}{\begin{itemize}}
\newcommand{\eit}{\end{itemize}}
\newcommand{\trm}[1]{\textrm{#1}}

\newcommand{\Mpc}{\trm{\Mpc}}
\newcommand{\yr}{\trm{\yr}}
\newcommand{\eV}{\trm{\eV}}

\newcommand{\vrad}{V}
\newcommand{\vd}{\sigma_0}

\allowdisplaybreaks

\begin{document}
\preprint{DESY-18-040}
\title{
 Core formation from self-heating dark matter
}

\author{Xiaoyong Chu}
\email{xiaoyong.chu@oeaw.ac.at }
\affiliation{Institute of High Energy Physics, Austrian Academy of Sciences, Nikolsdorfer Gasse 18, 1050 Vienna, Austria
}

\author{Camilo Garcia-Cely}
\email{camilo.garcia.cely@desy.be}
\affiliation{Deutsches Elektronen-Synchrotron DESY, Notkestrasse 85,
22607 Hamburg, Germany}

\hypersetup{
 pdftitle={Core formation from self-heating dark matter},
 pdfauthor={Xiaoyong Chu, Camilo Garcia-Cely}
}

\begin{abstract}

Cosmological simulations of the $\Lambda$CDM model suggest that the dark matter halos of dwarf galaxies  are  denser in their center than what observational data of such galaxies imply.   
In this letter, we propose a novel solution to this problem by invoking a certain class of dark matter  self-heating processes. As we will argue, such processes lead to the formation of dark matter cores at late times by considerably reducing the inner mass density of dwarf-sized halos. For deriving concrete results,  we focus on semi-annihilating dark matter scenarios and model the inner region of dark matter halos as a gravothermal  fluid.
An important aspect of this new solution is that the semi-annihilation effects  are much more prominent in dwarf-sized halos than in the more massive halos that host galaxies and clusters, even if the corresponding cross sections are the same.
Furthermore, the preferred parameter space for solving the small-scale problem suggests a thermal dark matter candidate with a mass below the GeV scale, which can be probed in dark matter direct and indirect detection experiments.

\end{abstract}

\maketitle
\section{Introduction}

The $\Lambda$CDM model is well accepted as the standard theory of astrophysical and cosmological phenomena.  Nonetheless,  there exists a large amount of evidence --mostly based on collisionless dark matter (DM) simulations without baryons-- which suggests that this model tends to overproduce the central  DM density of the small  halos  
in comparison with observations. Such a mass deficit in the inner halos, conventionally referred to as the ``core-vs-cusp''~\cite{Moore:1994yx, Flores:1994gz, McGaugh:1998tq} and ``too-big-to-fail''~\cite{BoylanKolchin:2011de,Garrison-Kimmel:2014vqa} problems, has been explained in the literature with various mechanisms that raise the entropy of the DM phase space. For instance, these include self-interacting DM (SIDM)~\cite{Spergel:1999mh}, or processes that directly heat up the DM particles in the center: slow decay of nearly-degenerate DM~\cite{Peter:2010jy, Medvedev:2013vsa, Wang:2014ina}, supernova-driven baryonic winds~\cite{Navarro:1996bv,Gelato:1998hb, Binney:2000zt, Gnedin:2001ec} or infalling baryonic clumps~\cite{ElZant:2001re, Weinberg:2001gm, Ahn:2004xt, Tonini:2006gwz}. See Refs.~\cite{DelPopolo:2016emo, Tulin:2017ara} for recent reviews that include a  comprehensive list of possible solutions. Here we propose a new  mechanism in which DM particle heats up itself, alleviating the aforementioned discrepancies between observations and cosmological simulations. 
For the purpose of illustration, we will focus on the case of semi-annihilating DM processes, which produce DM particles with large kinetic energies.

In order to study the impact of  DM self-heating processes on the halo density distribution, we will adopt the gravothermal fluid approximation~\cite{Heggie:2001re, 2008gady.book.....B}.  Such an approximation has been shown to describe reasonably well the features of weakly collisional systems at relatively long time scales~\cite{1970MNRAS:Larson, 1978PThPh:Hachisu, 1980MNRAS:Lynden}, 
For the sake of concreteness, several reasonable yet non-trivial simplifications have to be made.
Therefore, we emphasize that further N-body simulations will be necessary not only  to confirm our results but also to provide more precise predictions for observations. 

As will be explained, the mechanism proposed here has some novel features.  First of all, the heating-up effect only relies on DM particle properties and it is thus independent of baryons or environmental conditions.  Secondly, 
picobarn-scale interactions are enough to alleviate the small-scale problems  and there is no need to introduce strong self-scattering cross sections at the level of cm$^2$/g.  Besides, the physical effect has the desired velocity dependence to avoid stringent bounds from   cluster observations.

This paper is organized as follows. In Section~\ref{sec:fluid}, we first  show how shallow cores in DM halos can be generated from DM semi-annihilations. Then, we will quantitatively discuss their impact on halo dynamics in several examples. In Section ~\ref{sec:discussion}, we examine experimental constraints and discuss possible detection signatures of self-heating DM. We conclude in Section \ref{sec:end} by presenting a summary and an outlook of this work.  Appendices \ref{fluid:dynamics} and \ref{general:solutions} are respectively devoted to a summary of the gravothermal fluid approximation and derivations of analytical results.

\section{Core formation}
\label{sec:fluid}

\subsection{Absorption of heat in the fluid picture}

As stated above, we are interested in studying  halos in the presence of DM inelastic collisions that release energy. For concreteness, we consider the semi-annihilation process, $\text{DM}+\text{DM}\to \text{DM}+\phi$, where $\phi$  is a state much lighter than DM. 
In such a process, the final DM particle gains an amount of energy $\delta E$ per collision.
In the light of this, we model the halo evolution under the following assumptions:

\begin{itemize}
\item[(i)] We assume that the lighter state  $\phi$ quickly escapes  and does not have any effect on the halo subsequently.

\item[(ii)]  In general, DM  elastic collisions will  prevent a fraction of the energetic DM particles from leaving the halo, and also re-distribute the kinetic energy gained from semi-annihilations among all DM particles.
In average, such a re-distribution happens in the inner region of the halo.  Hence, the rate per unit mass at which the DM halo is heated will be characterized by  $ \rho \,\xi \delta E \langle \sigma v_\text{semi}\rangle /m^2 $,  where $\rho$ is the DM mass density, $m$ is the DM mass and $\langle \sigma v_\text{semi}\rangle $ is the inelastic cross section.  Here,   $\xi$ is a dimensionless coefficient determining the efficiency of the capture process, which is expected to be much smaller than one.

 Notice that  the  semi-annihilation cross section $\langle \sigma v_\text{semi}\rangle $ is typically velocity independent~\cite{Kamada:2017gfc, McDermott:2017vyk}. For example, that is the case if it is  induced by  contact interactions. With this in mind, we will assume that $\xi \delta E\langle \sigma v_\text{semi}\rangle $ is a fixed parameter unless otherwise stated.  

\item[(iii)]  Elastic scatterings or even  gravitational interactions thermalize the DM halo at the center~\cite{Balberg:2002ue}.  
This allows us to treat the DM particles as a gravothermal fluid.
See Appendix~\ref{fluid:dynamics} for details. 
Then,   the DM temperature and the entropy per unit  mass are  respectively given by
\begin{align}
T= m\vd^2 \, &&\text{and}&& s = \frac{1}{m} \log\left(\frac{\vd^3}{\rho}\right) + \text{const.}\,,
\label{eq:Tands_def}
\end{align}
where 
$\sigma_0$ is the DM velocity dispersion due to the thermalization.

Moreover, thermalization processes are assumed to be fast enough so that the temperature is almost uniform in the inner part of the halo.
Notice that this approximation is best suited for a radius smaller than both the DM mean free path associated with elastic scatterings and the gravitational scale height $r_G \equiv (\vd^2/4\pi G \rho)^{1/2}$~\cite{Balberg:2002ue}.

\end{itemize}

With the assumptions above, we can now specify how heat is absorbed. The change of heat per unit mass, $\delta q$, is determined by heat conduction and inelastic scatterings. 
 In our case, according to (iii), 
 in the inner part of the halo the temperature is approximately uniform and the conduction term is negligible.\footnote{
More precisely, the heat change per unit mass induced by conduction is given by
$({\delta q}/{ \delta t})_\text{conduction}  =\bm{\nabla}\cdot \left(\kappa \nabla T\right)/\rho \simeq 0\,.
\nonumber
\label{eq:dQ_dt}
$
In fact, the thermal conductivity $\kappa$ is also small (see Appendix~\ref{fluid:dynamics}).  
} Then, 
\begin{equation}
\frac{\delta q}{ \delta t}  \simeq \frac{\rho  {\cal J} T}{ m^2 }\langle \sigma v_\text{semi}\rangle\,,
\label{eq:dQ_dtn}
\end{equation}
with ${\cal J} \equiv \xi\, \delta E/T  = \xi/(4\sigma_0^2) $. 
This is just the energy absorption rate per mass, as explained above.   We will take ${\cal J}\gg1$ so that the energy absorption plays a major role in the halo dynamics.

\subsection{Outflow of DM particles}

\begin{figure*}[t]
\centering
{\large \hspace{1cm} Inverse power law\hspace{3cm}  NFW \hspace{3.5cm} Einasto ($\alpha=0.1$) }\\
\vspace{-.4cm}\hspace{0.1cm}
\includegraphics[height=0.26\textheight]{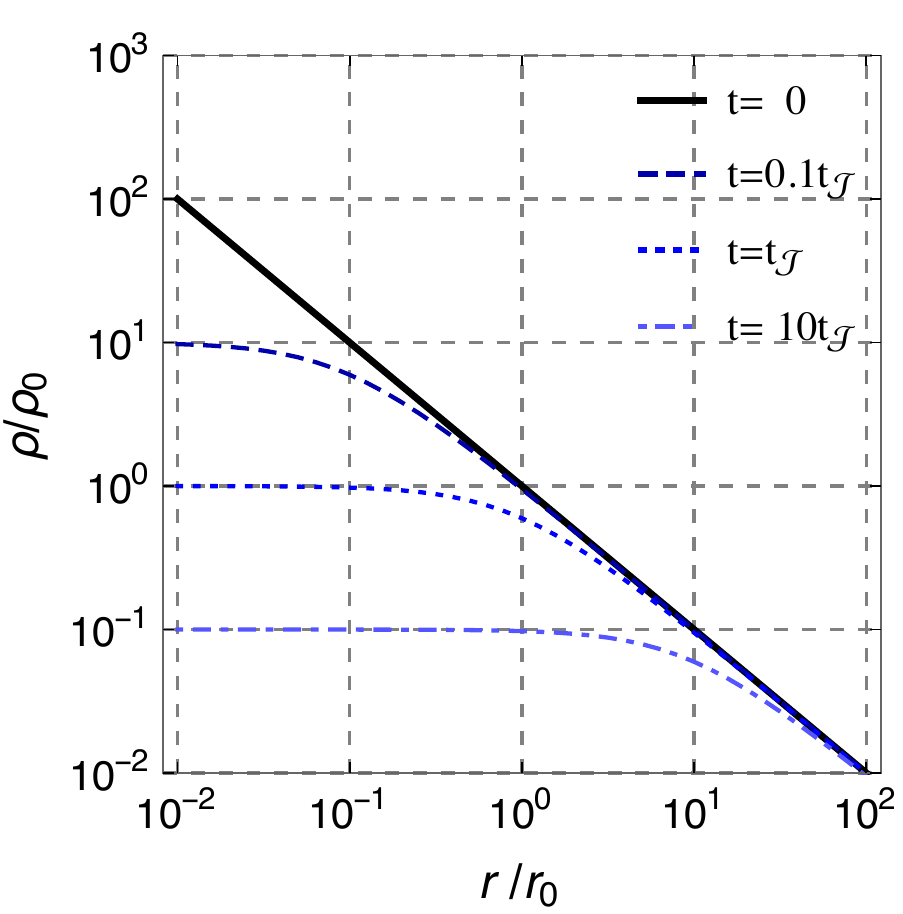}
\includegraphics[trim=0.66cm 0cm 0cm 0cm,clip,height=0.26\textheight]{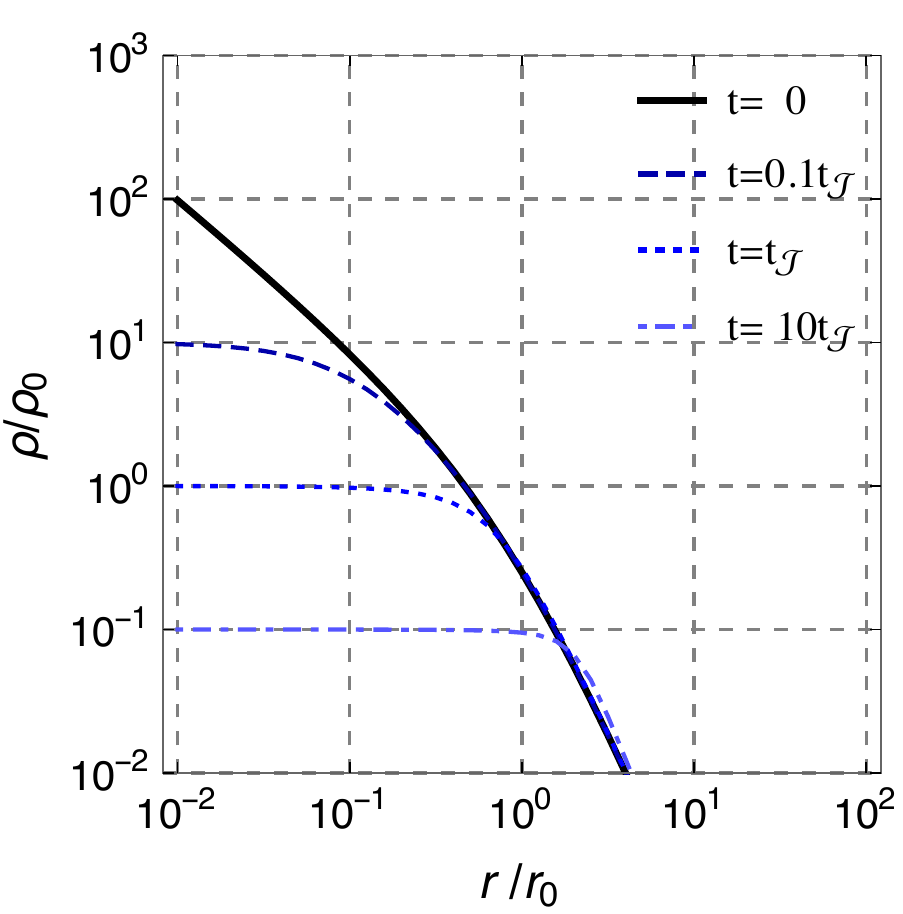}
\includegraphics[trim=0.66cm 0cm 0cm 0cm,clip,height=0.26\textheight]{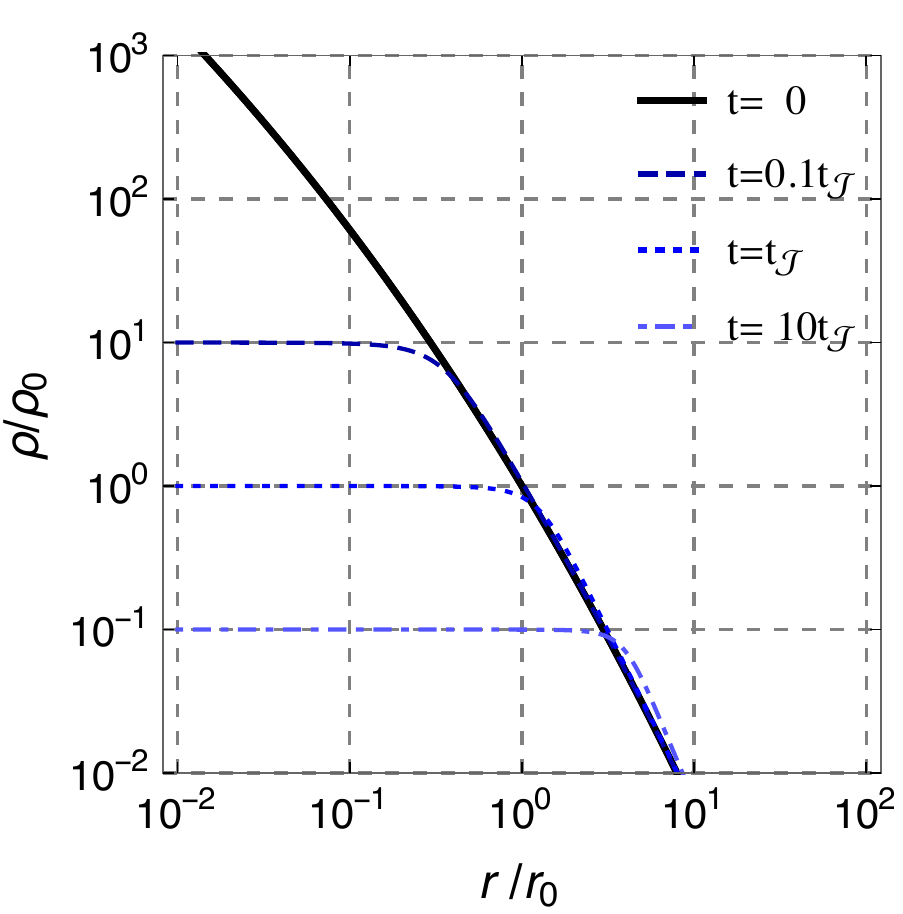}\\
\includegraphics[height=0.26\textheight]{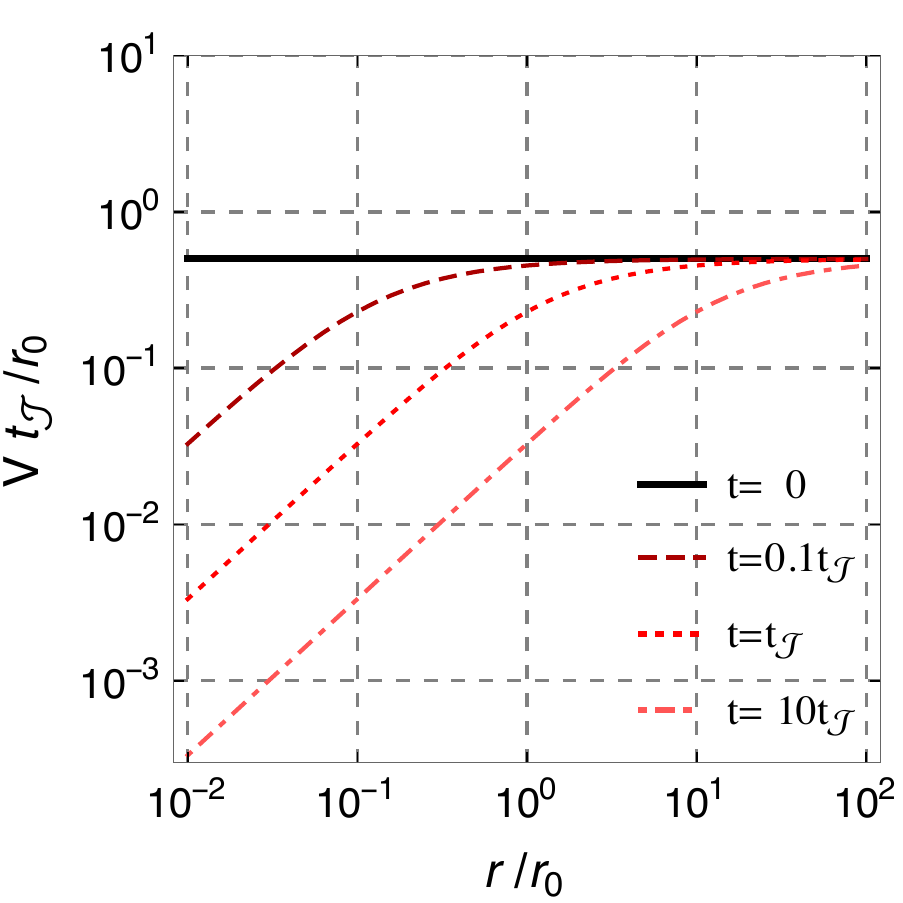}
\includegraphics[trim=0.66cm 0cm 0cm 0cm,clip,height=0.26\textheight]{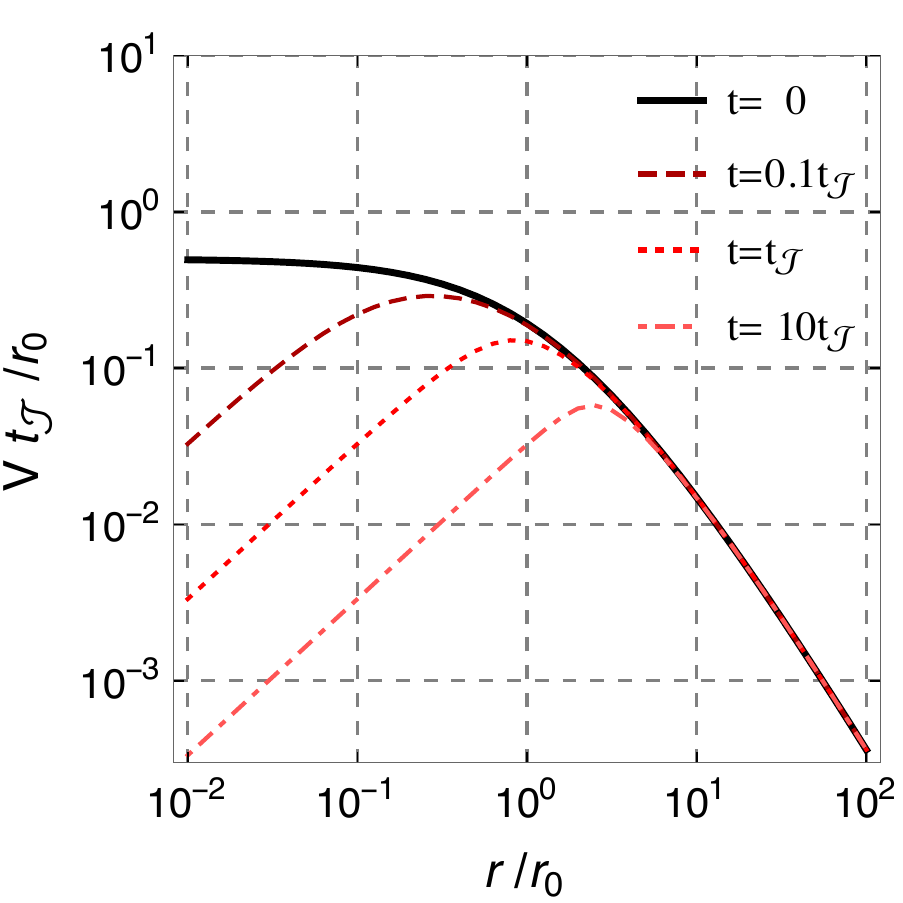}
\includegraphics[trim=0.66cm 0cm 0cm 0cm,clip,height=0.26\textheight]{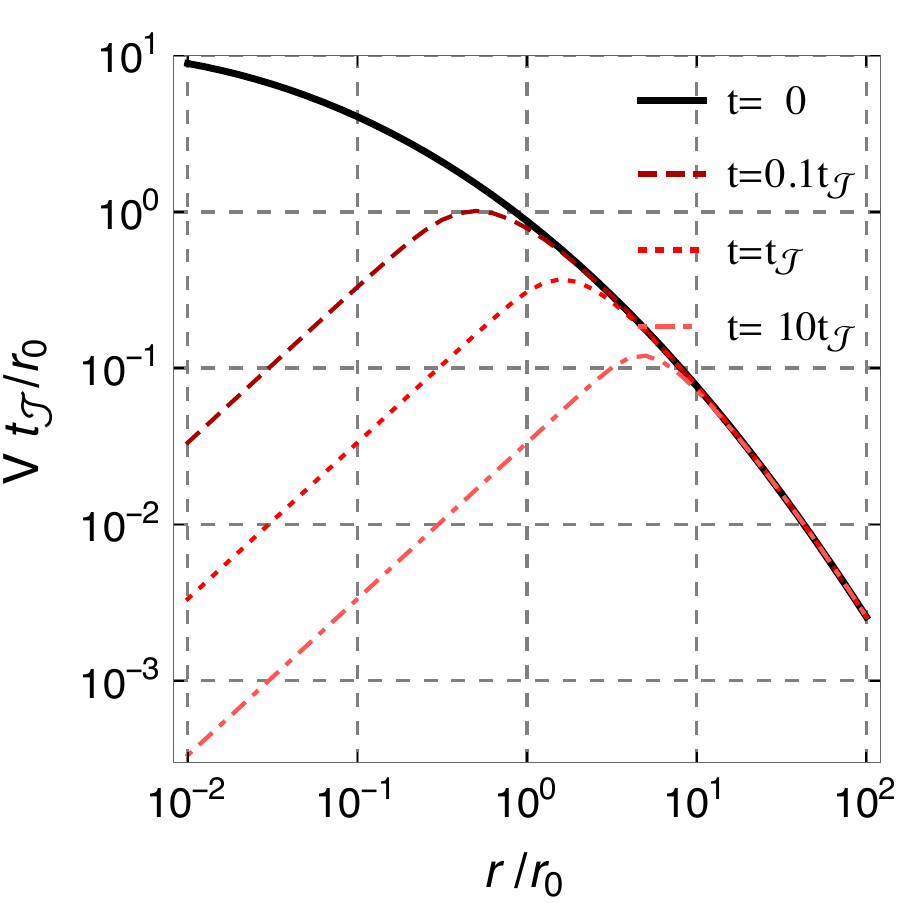}
\vspace{-.3cm}
\caption{\textit{Top panel:} The time evolution of the DM density in the presence of semi-annihilations. The initial condition (t=0) is shown in black and corresponds to the inverse power law\,(left),  the NFW\,(middle), and Einasto\,(right) profiles. The subsequent profiles at $t= 0.1t_{\cal J}, t_{\cal J}, 10 t_{\cal J}$  are shown as dashed, dotted, dash-dotted curves, respectively.  \textit{Bottom panel:} Same as the top panel but for the velocity evolution. In each panel quantities have been re-scaled with respect to $\rho_0$, $r_0/t_{\cal J}$ and $r_0$ (see Eq.~\eqref{DM:profiles}). 
}

\label{fig:fig1}
\end{figure*}

A slow release of heat increases the entropy, and according to Eq.~\eqref{eq:Tands_def}, it can raise the temperature and decrease the DM density.
A change in the latter is equivalent to a collective motion of particles, which can be described by a macroscopic velocity $\mathbf{V}$. We will assume that this process happens in quasi-static equilibrium, where $|\mathbf{V}| \ll \sigma_0$. 
In such a regime, Eq.~\eqref{eq:dQ_dtn} as well as the known relation between entropy change and absorbed heat,  $ds =\delta q/T$,
leads to
\begin{eqnarray}
\frac{d s}{d t} =\frac{\partial s}{\partial t} + \mathbf{\vrad} \cdot \nabla s
=
\frac{\rho}{ m^2 }\langle \sigma v_\text{semi}\rangle {\cal J}\, .
\label{eq:entropy_ev}
\end{eqnarray}

In addition, the evolution of DM mass density is also described by the continuity equation
\begin{equation}
\frac{\partial \rho}{\partial t}+\nabla \cdot (\rho \mathbf{\vrad}) = -\rho \frac{\delta N}{\delta t} \,.
\label{eq:continuity}
\end{equation}
Here, 
$\delta N/ \delta t \propto \rho \langle \sigma v_\text{semi}\rangle/m$, 
is the rate of change in the number of DM particles in the halo.
 The exact proportionality factor is a number in between one and two,  depending on the efficiency of the capture process. 
 
Furthermore, the pressure induced by the velocity dispersion, together with a subleading contribution from the collective motion of DM particles,  balances the gravitational force  on the DM, 
$
\vd^2 \nabla \rho + \rho\left(\partial \mathbf{\vrad}/{\partial t} + (\mathbf{\vrad} \cdot \nabla) \mathbf{\vrad} \right)   =- \rho\nabla \Phi
$.
Here $\Phi$ is the gravitational potential which is determined by the baryon and the DM density.  
The lack of knowledge of the baryon density precludes a detailed analysis of the gravitational dynamics of the halo.  
Nevertheless, Eqs.~(\ref{eq:Tands_def}), \eqref{eq:entropy_ev} and \eqref{eq:continuity}   already convey sufficient information about the  DM distribution in the center of the halo.
In fact, since  close to the center the velocity dispersion  is uniform,  they imply
\begin{eqnarray}
\nabla \cdot \mathbf{\vrad} +\frac{3}{\vd}\frac{d \vd}{dt} &=&-\frac{\delta N}{\delta t} +\frac{\rho}{ m }\langle \sigma v_\text{semi}\rangle {\cal J}\notag\\&\simeq& \frac{\rho}{ m }\langle \sigma v_\text{semi}\rangle {\cal J}\,.
\label{eq:master0}
\end{eqnarray}
In the last line, we have made use of the fact that we are focusing on processes such that ${\cal J}\gg1$,  as mentioned above. 
This is  a key equation describing the evolution of the halo, which  shows that the dynamic time scale of the halo  is determined by the inelastic interaction rate times ${\cal J}$,   instead of simply the interaction rate. 
Moreover, it implies that  
the positive entropy injection tends to either increase the velocity dispersion (i.e. conduction), or reduce the central mass density by creating a macroscopic outflow of particles (i.e. convection). 

For a realistic DM halo, convection is more efficient in transferring heat than thermal conduction~\cite{Balberg:2002ue}, as follows from the fact that the conduction term (see Appendix~\ref{fluid:dynamics}) is suppressed by both a small thermal conductivity and a nearly flat temperature profile in the inner halo.   Hence, to proceed further, from now on we neglect the time-derivative of the velocity dispersion and obtain 
\begin{equation}
\nabla \cdot \mathbf{\vrad} =\frac{\rho}{ m }\langle \sigma v_\text{semi}\rangle {\cal J},
 \hspace{10pt}\text{ or } \hspace{10pt}
\mathbf{V} =  \frac{\bar{\rho}}{3m}\langle \sigma v_\text{semi}\rangle {\cal J} \mathbf{r}\,,
\label{eq:V}
\end{equation}
where the last equation applies to only spherical halos. There,  $\bar{\rho}$ is the average DM density inside the sphere of radius $r$, i.e. its total mass divided by the volume.

We conclude that the outflow of DM particles is determined by the DM surface density ($\sim \bar{\rho} r$) and the strength of  energy absorption.  
Before discussing the consequences of this  for the formation of DM cores, 
let us apply  Eq.~\eqref{eq:V} to the quasi-static condition  $\vrad \ll \vd$. We obtain
\begin{equation}
\frac{\bar{\rho}\langle \sigma v_\text{semi}\rangle {\cal J}  / m}{H_0}\ll {3 \vd \over r H_0} \simeq  300  \left({\sigma_0 \over 5\,\text{km/s}}\right)\left({\text{kpc} \over r}\right)\,,\label{qsc:velocity}
\end{equation}
where $H_0$ is the expansion rate at present. On the other hand, in order for semi-annihilations to have observable effects,  the dynamic time scale of such processes must  not be much larger than the age of the Universe. That is,  the energy absorption rates of interest here should be comparable to the expansion rate. This is consistent with Eq.~\eqref{qsc:velocity}, so both conditions can be easily  satisfied for dwarf-sized halos, where $\sigma_0\sim {\cal O}(5) $\,km/s and typical distances are measured in kpc.

\subsection{Evolution of the inner density}

The macroscopic velocity given by Eq.~\eqref{eq:V}, together with the continuity equation,  allows one to calculate the DM density in the center of the halo.  In fact, analytical solutions are found if $\delta N/\delta t$ is neglected in Eq.~\eqref{eq:continuity}. We have  numerically verified that including such a term in the continuity equation does not alter the results as long as ${\cal J}\gg 1$.  

Quantitatively, $\partial \rho /\partial t+ \nabla \cdot (\rho \mathbf{\vrad})\simeq 0$ and Eq.~\eqref{eq:V} 
are a closed system of equations. As derived in Appendix~\ref{general:solutions},  the corresponding  solution is 
\begin{eqnarray}
\rho (t, r) &=& \rho(0,R)\left({1+\frac{\rho(0,R)\langle\sigma v_\text{semi}\rangle {\cal J} t}{m}  }\right)^{-1}\,,\label{density:evol}\\
\mathbf{V}(t,r)& =& \left(1-\left(\frac{R}{r}\right)^3\right) \frac{\mathbf{r}}{3t}\,,
\end{eqnarray}
where $R=R(t,r)$ is a function implicitly defined by 
\begin{eqnarray}
1&=&\left(\frac{R}{r}\right)^3 +\frac{3\,t}{ t_{\cal J}}\int^{R}_0\frac{\rho(0,r') r'^2 dr'}{\rho_0\, r^3}\,,
\label{eq:rcp}
\end{eqnarray}
with
\begin{equation}
t_{\cal J}\equiv\frac{m}{\rho_0 \langle \sigma v_\text{semi}\rangle {\cal J} } \,.
\label{eq:Jdef}
\end{equation}
Here we have introduced a time scale $t_{\cal J}$ with respect to any characteristic DM density $\rho_0$ of interest. The latter can be associated with the \emph{initial} density profile, as will be specified below. 
Note that at early times when $t\ll t_{\cal J}$, we  have $R(t,r) \simeq  r$, so the solution becomes $\rho (t, r) \simeq \rho(0,r)$, as expected. This also means that  at  such early times the elastic and inelastic collisions are effectively absent and therefore  $\rho(0,r)$ is essentially the same as that  in  collisionless DM models.

Cosmological simulations of halos composed of only collisionless DM suggest a sharp increase in the DM density at small radii. Motivated by results from those simulations (see e.g.~Refs.~\cite{Navarro:2008kc, Hayashi:2007uk}),  we consider the following cuspy profiles as initial conditions of Eqs.~\eqref{density:evol}-\eqref{eq:rcp}
\begin{equation}\label{DM:profiles}
\hspace{-0.3cm}\rho(0,r) = \left\{
\begin{array}{ll}
\frac{\rho_0 r_0}{r} & \text{(Inverse power law~\cite{Dubinski:1991bm}),}\\ [.3cm]
\frac{\rho_0}{(r/r_0) \left(1+r/r_0 \right)^2 } & \text{(NFW~\cite{1997ApJ...490..493N}),}\\  [.05cm]
 {\rho_0}e^{-{2\over \alpha }\left({r^\alpha \over r_0^\alpha } - 1 \right)}  & \text{(Einasto~\cite{Einasto:1965czb} with $\alpha=0.1$).}\\
\end{array}
\right.
\end{equation}
 Notice that the equation above explicitly defines $\rho_0$ and thus the value of $t_{\cal J}$ for each profile.

 The impact of the semi-annihilation process on the halo evolution is illustrated in Fig.~\ref{fig:fig1} for $t/t_{\cal J} = 0,\, 0.1,\, 1$, and $10$. For each profile, the top panel of the figure shows that the DM density in the inner halo is continuously decreasing due to the outward flow driven by entropy injection. Furthermore, the cusp in the DM density  that is initially present in each profile eventually  evolves into a core.  For the sake of illustration, the corresponding flow velocity for each profile is also shown in the bottom panel.  Notice also that, at large radii the DM density is essentially the same as its initial value, i.e. the one given by cosmological simulations. The semi-annihilation effects are thus only relevant in the inner regions. 

As a matter of fact, a generic prediction of this scenario is the formation of a growing DM core for $t > t_{\cal J}$, regardless of the initial condition. To see this, notice that Eq.~\eqref{eq:rcp} suggests that for large $t$ and small $r$ --when the second term on its R.H.S. dominates-- we  have $R(t,r)\ll r$ and hence   $ \mathbf{\vrad} (t,r) \simeq \mathbf{r}/{3t}$. In the previous three examples, this asymptotic behavior  can be directly seen  in the bottom panel of Fig.~\ref{fig:fig1}.  Since 
$\rho \simeq m \nabla \cdot \mathbf{\vrad}  /\langle \sigma v_\text{semi}\rangle  {\cal J} $, we conclude that the  DM density becomes uniform for sufficiently large times and small radii.  The region where this takes place defines a core with mass density\footnote{A similar core also appears for DM annihilations (e.g.~\cite{Gondolo:1999ef}). Nevertheless, in that case the core is expected to be much smaller  because  the corresponding core density is determined by the annihilation rate. } 
\begin{align}
\rho (t,r) \simeq \rho_c(t) =  \frac{m}{\langle \sigma v_\text{semi}\rangle {\cal J} t}\,.
\label{eq:asymp}
\end{align}
Notice that the initial condition does not enter in this equation. We have verified that such a relation holds for all the density profiles shown in the top panel of  Fig.~\ref{fig:fig1} when $t\gtrsim t_{\cal J}$ and $r\lesssim r_0$. This can be easily seen from the fact that $\rho_c (t)  = \rho_0  t_{\cal J}/t$.

It is worth pointing out that the core formation described by Eq.~\eqref{eq:asymp}  can be considered as a direct consequence of  the two basic arguments we have made in this section: $a)$ that the dynamical time scale is given by Eq.~\eqref{eq:Jdef}, and $b)$ that the entropy injection tends to modify the DM density (rather than the velocity dispersion). 
This suggests that  Eq.~\eqref{eq:asymp} should also apply to more general scenarios as long as these two arguments hold.
 One example is the case of decaying DM, where cosmological simulations show that similar heat absorption rates can solve the small-scale problems~\cite{Wang:2014ina}.

\section{Phenomenology of self-heating dark matter}
\label{sec:discussion}

As stated in the Introduction, our primary interest is the DM halos of dwarf galaxies. It is conceivable that they have already entered in the core-formation phase, that is, $t_\text{age}>t_{\cal J}$. Then, Eq.~\eqref{eq:asymp} leads to
\begin{equation}
\frac{\rho_c\langle \sigma v_\text{semi}\rangle {\cal J} t_\text{age} }{m} \simeq 1\,.
\label{eq:age}
\end{equation}

Observations suggest that dwarf-sized halos exhibit core densities with $\rho_c \sim 0.5\,\text{M}_\odot/\text{pc}^3$, velocity dispersions of the order $\sigma_0\sim 5$--$10$\,km/s and  are typically $10^{10}$\,years old. See Table 1 of Ref.~\cite{Walker:2009zp} for a comprehensive compilation of dwarf galaxies with their structural parameters.  Substituting these numbers into Eq.~\eqref{eq:age}, our model implies $ \sigma_\text{semi}  {\cal J}/m \sim \mathcal{O}(0.1) $\,cm$^2/g$.

Eq.~\eqref{eq:age} is reminiscent of the relation $\rho_c \langle \sigma v_\text{elastic}\rangle t_\text{age}/m \simeq 1$, arising for the core formation process in the SIDM scenario~\cite{Kaplinghat:2015aga}.  In the latter case, the DM entropy increases in the inner halo because of  elastic scatterings, which are characterized by $\langle\sigma v_\text{elastic}\rangle/m$. In contrast, the entropy change in our model is driven by the energy-absorption cross section per unit  mass, $ \langle\sigma v_\text{semi}\rangle {\cal J}/m$. 
Quantitatively, the corresponding velocity-averaged cross section is
\begin{eqnarray}
\langle\sigma v_\text{semi}\rangle &\sim&\frac{\unit[10^{-29}]{cm^3/s}}{\xi}  
\left(\frac{m}{\unit[100]{MeV}}\right) \left(\frac{\sigma_0}{\unit[5]{km/s}}\right)^2 \nonumber \\
&&
\times
\left(\frac{ 0.5 \text{M}_\odot/\text{pc}^3}{\rho_c}\right)
\left(\frac{\unit[10^{10}]{years}}{t_\text{age}}\right)\,.
\label{eq:sigmav}
\end{eqnarray} 
Hence, roughly speaking, the formation of a shallow core can be explained by cross section values below a few picobarns, 
much smaller than those needed in the SIDM case (typically $\sigma_\text{elastic}/m \sim  \text{mb}/\text{MeV}$). 

\begin{figure}[t]
\vspace{-.5cm}
\centering
\includegraphics[width=0.5\textwidth]{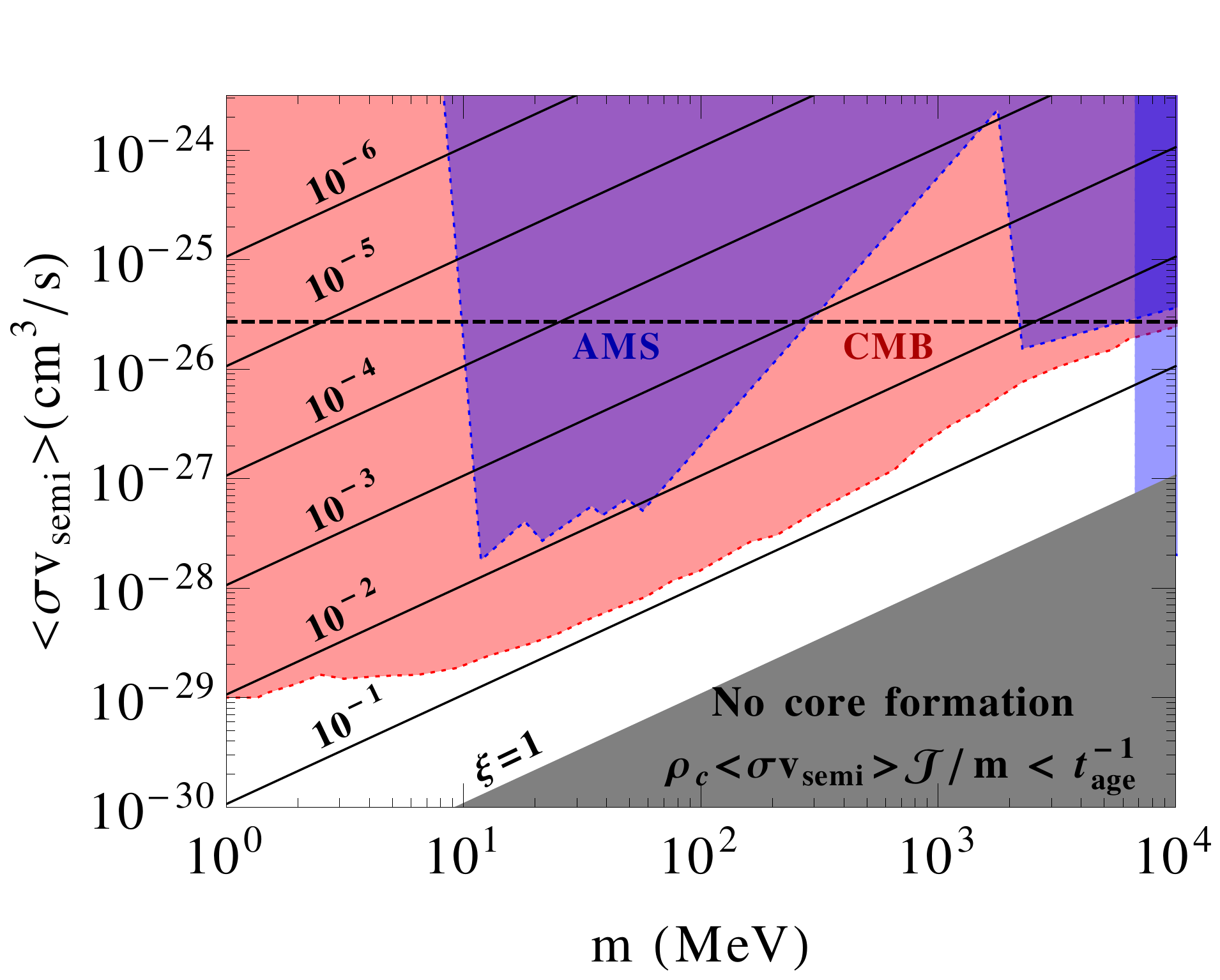}
\vspace{-.5cm}
\caption{Semi-annihilation cross sections as a function of DM mass. The values of $\langle \sigma v_\text{semi}\rangle$ preferred by the observed core in dwarf-sized halos are given by black solid lines for various capture efficiencies $\xi$. 
If the light particle $\phi$ interacts \emph{electromagnetically}, the CMB limits from Ref.~\cite{Slatyer:2015jla} exclude the red shaded region.  AMS bounds on semi-annihilations into states producing positrons are given in the blue shaded region\cite{Aguilar:2014mma, Boudaud:2016mos,Bergstrom:2013jra}. Above the dashed line, a DM production mechanism other than thermal freeze-out is required. 
}
\label{fig:fig2}
\end{figure}

For the more massive halos that host galaxies and clusters,   the velocity dispersions are of the order $100-1000$\,km/s, so their values of ${\cal J}$ are only $10^{-2}$ to $10^{-4}$ of those of dwarf-sized halos.  According to Eq.~\eqref{eq:Jdef}, the dynamical time scale $t_{\mathcal J}$ becomes significantly larger in massive halos, and as a result  DM semi-annihilation does not play an observable role there. That is, observational data from clusters does not constrain our scenario.\footnote{  While core formation does not happen in cluster-sized halos, the latter might be affected by the fact that they grow hierarchically from accumulation of smaller halos where semi-annihilations can be relevant~\cite{BoylanKolchin:2003sf}. The study of such an effect requires a cosmological simulation, which lies beyond the scope of this work.   }
This is another difference in comparison with the SIDM scenario. The latter -- without semi-annihilations -- is disfavored if the DM self-scatterings are induced  by a contact interaction. In that case, $\sigma_\text{elastic}/m$ is velocity independent and the corresponding value needed to address the small-scale problems in  dwarf galaxies  generates too shallow cores in very massive halos, in apparent contraction with observations of galaxy clusters~\cite{Kaplinghat:2015aga, Elbert:2016dbb}.\footnote{ The most common alternative to this is to introduce a light mediator in order to obtain a velocity-dependent $\sigma_\text{elastic}/m$ so that  the cluster bounds from above are avoided. Nonetheless, if the light mediator decays visibly,  it also induces a Sommerfeld enhanced signal that spoils CMB observations in most cases~\cite{Bringmann:2016din}.} 
Furthermore, in contrast to the SIDM case, the evaporation of the small subhalos moving inside cluster-sized halos does not constrain our scenario. This is because the evaporation rate is given by
\begin{equation}
\Gamma_\text{ev}=
 \frac{\rho (1-\xi)}{m_\text{DM}}\langle \sigma v_\text{semi} \rangle \ll 
\frac{\rho}{m_\text{DM}}\langle \sigma v_\text{semi} \rangle  {\cal J}\,.
\end{equation}
Then, according to Eq.~\eqref{eq:age},  the evaporation time scale is much larger than the age of the halo, making  evaporation constraints negligible.

There are two important implications of Eq.~\eqref{eq:sigmav}. On the one hand, 
in the case of a velocity-independent $\langle \sigma v_\text{semi}\rangle $, the abundance of DM can be thermally produced in the early Universe  (for $m\lesssim\unit[10]{GeV}$). Depending on $\xi$,  the thermal freeze-out of the semi-annihilations could solely give rise to the observed relic density. Nevertheless, other processes could also play the dominant role at freeze-out. This is plausible given the fact that semi-annihilations such as $\text{DM}+\text{DM}\to\text{DM}+\phi$ typically take place together with $\text{DM}+\text{DM}\to \phi +\phi$ or  with $3\to2$ processes~\cite{Hochberg:2014dra, Bernal:2015bla}. We would like to remark that if the semi-annihilation process is velocity suppressed, the DM abundance can not be generated from a thermal freeze-out.   Despite all this,   $\langle \sigma v_\text{semi}\rangle$ fairly different from $\unit[3\times10^{-26}]{cm^3/s}$ is also conceivable   
if a production mechanism other than thermal freeze-out is invoked, as has been done for SIDM (see e.g.~\cite{Bernal:2015ova}).

On the other hand, Eq.~\eqref{eq:sigmav} also suggests that this scenario can be tested by means of the indirect searches of the light particle $\phi$ produced in the semi-annihilation process. We first consider the case in which $\phi$ interacts electromagnetically.  Then, indirect searches from astrophysical objects are typically not competitive with respect to cosmic microwave background (CMB) bounds for sub-GeV DM. This is because of the so-called \emph{MeV gap},  where current indirect detection experiments are not sensitive enough. In fact, currently the only relevant astrophysical constraint comes from considering DM semi-annihilations into a light particle that subsequently produces positrons. In that case, results from AMS antimatter searches exclude cross section larger than $\unit[10^{-28}]{cm^3/s}$ for DM  heavier than several GeV or in the mass range in between 10 and 70\,MeV~\cite{Aguilar:2014mma, Boudaud:2016mos}.

At the same time, DM semi-annihilations during the dark ages -- the time between recombination and reionization, when CMB photons travel freely in the Universe -- could modify the CMB temperature spectrum and its polarization  modes via the electromagnetic interactions of $\phi$.  
Consequently, the precise measurements from  \textit{Planck}~\cite{Ade:2015xua} set stringent constraints on $\langle \sigma v_\text{semi}\rangle$ at high redshifts. We use the results from Ref.~\cite{Slatyer:2015jla} to constrain our scenario after adapting them for the s-wave semi-annihilations considered here. 

The aforementioned AMS and CMB  exclusions as well as the regions where Eq.~\eqref{eq:age}  is fulfilled are shown in Fig.~\ref{fig:fig2}. These  correspond to the blue and red regions, and the area on the bottom-right gray corner, respectively.  We have taken  $\sigma_0 =\unit[5]{km/s}$, $\rho_c = \unit[0.5]{\text{M}_\odot/\text{pc}^3}$ and $t_\text{age} =\unit[10^{10}]{years}$~\cite{Walker:2009zp}.  Note that the remaining white region will be  probed by future CMB--S4 experiments~\cite{Madhavacheril:2013cna}. Moreover, if the DM (or the light particle $\phi$) is in chemical equilibrium with the baryon-photon plasma, $m \lesssim \mathcal{O}(10)$\,MeV can be excluded by the bound on $N_\text{eff}$ derived from \emph{Planck}~\cite{Boehm:2013jpa}. Nevertheless, since such exclusion is model dependent~\cite{Green:2017ybv, Berlin:2017ftj}, we do not depict it in the plot. 

Fig.~\ref{fig:fig2} suggests two experimentally allowed  possibilities. One of them,  corresponding to the white region, is that the capture efficiency is relatively large with $\xi \gtrsim 0.03$. The other one is that the CMB/AMS bounds in the colored regions do not apply.  One simple realization of  the first possibility  corresponds to situations where the elastic collisions are frequent enough to efficiently trap the energetic particles in the DM halo. Therefore, the elastic scattering processes must be sufficiently strong, with $\sigma_\text{elastic}/m$ not far from ${\mathcal O}(0.1)$\,cm$^2$/g. 
Such a realization is in fact a combination of the semi-annihilating DM and the standard SIDM scenario with contact interactions, in which $\langle \sigma v_\text{semi}\rangle$ and $\sigma_\text{elastic}$ are both velocity independent. Determining whether such a realization is still compatible with the cluster observations discussed above requires a more dedicated study that includes a numerical fit of N-body simulations of the full model to the observational data. Nonetheless, it is clear that our scenario with Eq.~\eqref{eq:sigmav} can potentially allow for  SIDM models with constant $\sigma_\text{elastic}/m\lesssim \unit[0.2]{cm^2/g}$~\cite{Harvey:2015hha, Elbert:2016dbb} while addressing the small-scale problems.

The other possibility is that the CMB/AMS bounds do not apply, which opens up the parameter space.  
 This corresponds to the cases where the DM semi-annihilation is  velocity suppressed or only produces neutrinos or other invisible particles. 
For instance, if DM semi-annihilates into  only neutrinos,
relevant indirect detection bounds are well above the thermal cross section (see e.g.~Ref.~\cite{Beacom:2006tt}), making our scenario effectively unconstrained. Likewise,  the light particle $\phi$ involved in the semi-annihilation process can also be a sub-leading DM component or dark radiation~\cite{McDermott:2017vyk},  in which case  the  parameter region shown in Fig.~\ref{fig:fig2} is  also unconstrained.

A final remark is in order before the end of this section. 
Notice that, since the semi-annihilation products
are relativistic, once they reach Earth, they 
can transfer detectable recoil energies to the  targets of direct detection experiments  via elastic collisions.
This leads to signatures similar to those of boosted DM scenarios~\cite{Agashe:2014yua}.  
Nevertheless, since the mechanism proposed here does not rely on the couplings between DM and the Standard Model particles, these associated constraints will be discussed elsewhere.

\section{Summary and Outlook}

\label{sec:end}

In this work, we have investigated self-heating DM processes in the center of  DM halos  as  a plausible explanation for the small-scale problems of the $\Lambda$CDM model.  For doing so,  we have focused  on the particular case of semi-annihilating DM and have modeled the central region of the halos as a  fluid meeting the criteria (i)-(iii) of Section~\ref{sec:fluid}. Under this set of simplifying assumptions,  we have shown that DM mass distributions exhibiting a cusp at early  times eventually form a shallower core.  We have argued analytically that this phenomenon is the result of an outflow of DM particles induced by the heat injection.  Our main results are also numerically illustrated in Fig~\ref{fig:fig1} for cases in which the initial DM density is described by several well-motivated cuspy profiles.

Subsequently, we have discussed the phenomenological implications of our hypothesis. First, our DM candidate can be thermally produced  if its mass is lighter than $\sim$\,10\,GeV. Furthermore, 
the scenario proposed in this work can be probed by  the indirect searches of  the light particles produced in the semi-annihilation processes, as illustrated in Fig.~\ref{fig:fig2}. In addition, it can be tested in direct detection experiments by observing recoils induced by the boosted particles from such processes.

The scenario proposed here
has certain advantages in comparison with SIDM models. Although
both of them increase the entropy in the inner halo to form a shallow core, the relevant dynamics in our case is determined by the ratio of the  energy absorption to the average kinetic energy. Accordingly,  our scenario has a much stronger effect in  dwarf galaxies than in clusters of galaxies, even if the relevant cross sections are the same in both objects.  In this sense, our proposal is an alternative to SIDM models with light mediators, which predict similar effects but are severely constrained in most cases.

On the other hand, the model presented here predicts the same galaxy rotation curves as SIDM because both give rise to the same type of density profiles, i.e. an inner core and an outer profile described by collisionless DM. In particular, the correlation  between the concentration parameter and the halo mass observed in cosmological simulations  is expected to hold in this scenario as well (see Ref.~\cite{Bondarenko:2017rfu}).  This suggests that the SIDM solution to the diversity issue proposed in Ref.~\cite{Kamada:2016euw}, i.e. introducing baryonic contraction, also applies to our scenario, although the quantitative effect may vary. A conclusive statement nevertheless requires more delicate treatments of baryonic effects. We leave this for future work.

Our modeling of the DM halo is a dramatic simplification of its actual evolution in the presence of processes releasing heat. For instance,  a more precise study based on the gravothermal fluid approximation can be made by considering conduction effects and a time-dependent velocity dispersion. Furthermore, a fully quantitative study requires N-body simulations of the inelastic and the capture processes. In spite of that, in all those cases, we expect similar results because the basic picture of the proposed mechanism is very clear: there is a DM process injecting energy in the central parts of the halo which creates an outflow of particles that diminishes its central abundance. 
Moreover, although the idea proposed in this work is illustrated in the context of semi-annihilating DM,  it can be generalized to other particle-physics models where DM heats up itself.

\section*{Acknowledgments}
We would like to thank Nicol\'as Bernal for his participation in the initial stages of this project. We also thank Kai Schmidt-Hoberg for discussions.     
 X.C. is supported by the ‘New Frontiers’ program of the Austrian Academy of Sciences. C.G.C. is supported by the German Science Foundation (DFG) under the Collaborative Research Center (SFB) 676 Particles, Strings and  the Early Universe as well as the ERC Starting Grant ‘NewAve’ (638528).
\appendix

\section{The fluid approximation}
\label{fluid:dynamics}
The evolution of the DM gravothermal fluid considered here can be described by the following set of equations~\cite{2008gady.book.....B}:
\begin{align}
\frac{\partial \rho}{\partial t} +\nabla \cdot (\rho \mathbf{\vrad}) &=- \frac{\rho^2}{m} \langle\sigma v_\text{semi}\rangle \,, \label{eq:density_evolution}\\
\rho\left(\frac{\partial \mathbf{\vrad}}{\partial t} + (\nabla \mathbf{\vrad} \cdot \nabla) \mathbf{\vrad} \right)&=- \rho\nabla \Phi -\nabla p \,,\\
\nabla^2 \Phi &= 4\pi G (\rho_b+\rho)\,,
\end{align}
where the DM pressure is determined by  $p = \rho T/m$.  Moreover, the expressions for the DM temperature $T$ and entropy per unit mass $s$ have been given in the main text, with the latter changing according to
\begin{align}
T\left(\frac{\partial s}{\partial t}+\mathbf{\vrad}\cdot \nabla s \right) &= {\delta q\over \delta t}\bigg|_\text{conduction}+{\delta q\over \delta t}\bigg|_\text{absorption}   \,. 
\label{eq:entropy_evolution}
\end{align}
Since $s$ is defined as entropy per unit mass, 
DM number reduction due to semi-annihilations also contributes to the change of $s$. In fact, this adds  a new term to the R.H.S. of Eq.~\eqref{eq:entropy_evolution}, which is given by ${(5\rho  /2m^2}) \langle\sigma v_\text{semi}\rangle T $. However, this term is
much smaller than  the heat absorption per unit mass
\begin{align}
	  {\delta q \over \delta t}\bigg|_\text{absorption} &=  {\rho  \over m^2} \langle\sigma v_\text{semi}\rangle \left( \xi   \delta E   \right)\, .
\end{align}
This is because the heating rate is much larger than semi-annihilation rate (or equivalently, ${\cal J}\equiv \xi\delta E/T \gg 1$).  
Therefore, the number-reduction contribution  can be absorbed in ${\cal J}$. In addition, 
\begin{align}
	  {\delta q \over \delta t}\bigg|_\text{conduction} &  =\frac{\bm{\nabla}\cdot \left(\kappa \nabla T\right)}{\rho}  ,
\label{entropy:deriv}
\end{align}
where the thermal conductivity is given by $\kappa^{-1} = \kappa_\text{scat}^{-1} +  \kappa_G^{-1}$~\cite{1968MNRAS:Lynden} with
\begin{align}
\kappa_\text{scat}
  \simeq  \frac{\sigma_0^2}{\langle\sigma v_\text{elastic}\rangle }\,,& &
 \kappa_G \simeq \frac{ \vd^2 \langle\sigma v_\text{elastic}\rangle \rho}{ 4\pi G m^2}\,.
\end{align}
Note that $\kappa_G \le  \kappa_\text{scat}$ as a result of $\rho \langle\sigma v_\text{elastic}\rangle /m < H_0 <  (4\pi G \rho)^{1/2}$ for dwarf-sized halos. This implies that $\kappa \simeq \kappa_G$. 

To sum up, the evolution of the inner halo caused by DM semi-annihilation can be obtained by solving these equations (\ref{eq:density_evolution}-\ref{entropy:deriv}) with certain initial conditions.

Naively one might expect that this system has static solutions where both $\rho$ and $s$ do not change with time, and thus $V\sim 0$. It turns out that this is not true because ${\cal J}\gg 1 $ and $\kappa$ is relatively small.  As an example, take the case of $\rho_b \ll \rho$,  the fact that $\kappa \simeq \kappa_G$ then allows to estimate the conduction rate as
\begin{equation}
	\left| {1 \over \rho }\bm{\nabla}\cdot \left(\kappa \nabla T\right) \right| \le 4\pi G m \kappa_G \simeq \frac{\rho}{m^2 }\langle \sigma v_\text{elastic}\rangle   m \vd^2\, .
\end{equation}
This is always smaller than the heat absorption rate, as given by Eq.~\eqref{entropy:deriv}, as long as $\langle \sigma v_\text{semi}\rangle  {\cal J} $ dominates over  $\langle \sigma v_\text{elastic}\rangle$. That is, Eq.~\eqref{eq:entropy_evolution} cannot be satisfied in a static solution, which has also been confirmed numerically.
In practice, this means that convection, characterized by the macroscopic velocity $\mathbf{\vrad}$,    plays the dominant role in transferring heat in  the halos considered here. 

\section{Analytical solutions}
\label{general:solutions}
In this appendix, we discuss the analytical solution of the system of equations $\partial \rho /\partial t+ \nabla \cdot (\rho \mathbf{\vrad})=0$ and $\nabla \cdot \mathbf{\vrad} =\rho\langle \sigma v_\text{semi}\rangle {\cal J}/m$. For doing this, as stated in assumption (ii) of Section~\ref{sec:fluid}, we assume that $\langle \sigma v_\text{semi}\rangle {\cal J}$ is constant. 
Then, the vector $\partial \mathbf{\vrad}/\partial t+(\nabla \cdot \mathbf{\vrad}) \mathbf{\vrad} $ is divergenceless. In fact, since $\mathbf{\vrad}$ points in the radial direction, that can only happen if $\partial \mathbf{\vrad}/\partial t+(\nabla \cdot \mathbf{\vrad}) \mathbf{\vrad} $  is exactly zero. We can write that as
\begin{equation}
\frac{1}{M}\frac{\partial M}{\partial t}+ \frac{\langle \sigma v_\text{semi}\rangle {\cal J}}{m}\frac{\partial M}{\partial {\cal V}} =0\,.
\end{equation}
For this, we used Eq.~\eqref{eq:V}, with $\bar{\rho} = M/{\cal V}$, where $ {\cal V}={4\pi r^3}/{3} $ is the volume containing the mass $M$. If we take $M$ and $t$ as the independent variables, the previous equations can be cast in a simpler form. Namely, $\left({\partial {\cal V}}/{\partial t}\right)_M =  {\langle \sigma v_\text{semi}\rangle {\cal J}} M/{m}$. This means that the volume containing a fixed amount of mass increases according to 
\begin{equation}
{\cal V}(t,M)={\cal V}(0,M)+\frac{\langle \sigma v_\text{semi}\rangle {\cal J}}{m} Mt \,. 
\label{eq:int1}
\end{equation}
Hence, the function $R=R(t,r)$ defined by means of ${\cal V}(0,M)= 4\pi R^3/3$ satisfies 
\begin{equation}
1=\left(\frac{R}{r}\right)^3 + \frac{3 \langle \sigma v_\text{semi}\rangle {\cal J} t}{m r^3}\int^{R}_0 \rho(0,r') r'^2 dr'\,.
\label{eq:rc}
\end{equation}
Furthermore, using Eqs.~\eqref{eq:V} and \eqref{eq:int1} we find that 
\begin{eqnarray}
\mathbf{V} = \left(1-\left(\frac{R}{r}\right)^3\right) \frac{\mathbf{r}}{3t}\,.
\label{eq:VV}
\end{eqnarray}
The divergence of this expression yields 
\begin{eqnarray}
\rho (t, r) = \rho(0,R)\left({1+\frac{\rho(0,R)\langle\sigma v_\text{semi}\rangle {\cal J} t}{m}  }\right)^{-1}\,.
\end{eqnarray}
In summary, a knowledge of the initial density profile, $\rho(0,r)$, allows to calculate the function $R$ by means of Eq.~\eqref{eq:rc}, which --according to the last two equations-- determines the velocity and the DM density at all times.

\bibliographystyle{utcaps_mod}
\bibliography{ref}

\end{document}